\documentclass[journal]{IEEEtran}
\usepackage[utf8]{inputenc}
\author{hak369 }
\date{November 2020}
\usepackage{verbatim}
\usepackage{optidef}
\usepackage{amssymb}
\usepackage[tableposition=top]{caption}
\usepackage{diagbox}

\let\llncssubparagraph\subparagraph
\let\subparagraph\paragraph
\usepackage[compact]{titlesec}
\let\subparagraph\llncssubparagraph

\usepackage{cite}
\ifCLASSINFOpdf
  \usepackage[pdftex]{graphicx}
  
  \DeclareGraphicsExtensions{.pdf,.jpeg,.png}
\else
 
  \usepackage[dvips]{graphicx}

 \DeclareGraphicsExtensions{.eps}
\fi
\usepackage{amsmath}
\usepackage{comment}
\usepackage{amssymb}
\usepackage{multirow}
\usepackage{array}
\usepackage{enumitem}
\newcolumntype{P}[1]{>{\centering\arraybackslash}p{#1}}
\newcolumntype{M}[1]{>{\centering\arraybackslash}m{#1}}
\titlespacing*{\section}
{0pt}{0pt}{0pt}
\titlespacing*{\subsection}
{0pt}{0pt}{0pt}
\titlespacing*{\subsubsection}
{0pt}{0pt}{0pt}
\usepackage[rgb,svgnames]{xcolor}

\usepackage[markup=underlined, authormarkup=default]{changes}
\definechangesauthor[color=Blue]{CA}

\begin{document}

\title{Inequitable Access to EV Charging Infrastructure}

\author{Hafiz~Anwar~Ullah~Khan,~\IEEEmembership{Student Member,~IEEE},  Sara Price, Charalampos Avraam, and Yury Dvorkin,~\IEEEmembership{Member,~IEEE}
\vspace{-25pt}
}

\maketitle

\begin{abstract}
Access to and affordability of electric vehicle (EV) charging infrastructure are the two prominent barriers for EV adoption. While major efforts are underway in the United States to roll-out public EV charging infrastructure, persistent social disparities in EV adoption call for interventions.
In this paper, we analyze the existing EV charging infrastructure across New York City (NYC) to identify such socio-demographic and transportation features that correlate with the current distribution of EV charging stations. Our results demonstrate that population density is not correlated with the density of EV chargers, hindering New York's EV adoption and decarbonization goals. On the contrary, the distribution of EV charging stations is heavily skewed against low--income, Black--identifying, and disinvested neighborhoods in NYC, however, positively correlated to presence of highways in a zip code. The results underscore the need for policy frameworks that incorporate equity and justice in the roll-out of EV charging infrastructure.
\end{abstract}

\IEEEpeerreviewmaketitle

\section{Introduction} \label{intro}
Transportation is a major source of greenhouse gas emissions and air pollution, with emissions from light-duty vehicles constituting its major share. For example, the light-duty vehicles in New York City (NYC) emit 80\% of the city's total transportation emissions \cite{DOT}. Electrified transportation is one of the critical aspects of the global trend towards decarbonization. With light-duty electric vehicle (EV) prices rapidly declining to as low as \$18,875 \cite{price} (after United States (US) federal tax credits and state rebates \cite{NYSERDA}) and their ranges expanding to 400 miles \cite{range}, it is anticipated that access to charging infrastructure will become the most prominent adoption barrier for EVs. The significance of the availability and affordability of charging infrastructure for adopting EVs is difficult to understate. From a planning perspective, insufficient EV charging infrastructure manifests itself in suppressed EV demand, discouraging private sector investments in EV charging. Thus, public investments and policy incentives are required for seeding EV charging infrastructure market \cite{DOT}. Similarly, from the consumers' perspective, the availability of public EV charging is an important factor in decisions for EV purchases in the US \cite{review}. For instance, a 2017 online survey of US EV owners found that public charging and access to fast charging were viewed as top criteria when buying an EV \cite{ABB}. In line with \cite{ABB}, the survey in \cite{Nissan} determines that a “\textit{lack of charging facilities in my area}” was the third-ranked reason for not purchasing an EV and a “\textit{lack of quick charging stations}” the fourth. Thus, access to EV charging infrastructure shares a symbiotic relationship with EV adoption, and subsequently with the global decarbonization efforts. Therefore, the problem of access to and affordability of public EV charging infrastructure is critical for all stakeholders in EV roll-out, including investor-owned charging companies, electric power utilities, consumers, and regulators. 

To alleviate the accessibility and affordability barriers in charging infrastructure and spur investments in EV adoption, the US government proposes an ambitious \$7.5 billion plan for installing 500,000 EV charging stations across the US by 2030 \cite{CNBC, Reuters}. Similarly, many states in the US have embarked on programs for deploying public EV charging infrastructure. For example, Con Edison, an electric power utility in NYC, will install the first 100 curbside EV charging ports in 2021, under the New York Reforming the Energy Vision (NYREV) program \cite{ConEd_Curbside}. In the same year, Kansas City, Missouri, plans to install 30 to 60 EV chargers under its Right-Of-Way project \cite{Kansas}. Other curbside EV charging station projects are underway in California, Washington, New Jersey, and Ohio \cite{NYSERDA_CURB, Samrat}. However, these deployments are limited in size and do not close the constantly growing charging-capacity gap \cite{Electrek}. Moreover, accessibility of EV charging infrastructure is redundant if it is not affordable. Therefore, any roll-out of public EV charging infrastructure should ensure that its costs and benefits are equitably distributed in the society.  Currently, the burden of this roll-out disproportionately affects low-- and middle--income, underrepresented, and disadvantaged communities, exacerbating the already present racial, financial, and cumulative social disparities in EV adoption \cite{Brockway}. Hence, as governments reconcile decarbonization efforts and policy with environmental justice, it is imperative that equitable access to EVs, catalyzed by EV charging infrastructure, is ensured \cite{WhiteHouse}.   


Current literature addresses some aspects of equitable transition towards electrified transportation and disparities in access to EV charging infrastructure across race and income. For instance, Cheyne \textit{et al.} conclude that disadvantaged and minority communities are disproportionately affected by environmental and transportation injustice \cite{Cheyne}. Hardman \textit{et al.} extend these results by showing that the current EV charging infrastructure is not equitably dispersed and EV incentives do not support low--income buyers. This skews the EV buying power towards predominantly white, male, high-income, and educated households \cite{Hardman}. Similarly, lack of access to EV charging infrastructure near multi-unit housing units (mostly inhabited by low-income communities) is a key barrier in EV adoption \cite{Canepa}. A census block group--level analysis in California shows that Black-- and Hispanic--majority neighborhoods have lower access to public EV charging infrastructure \cite{Hsu}. Brockway \textit{et al.} investigate the effects of grid limits on the growth and adoption of Distributed Energy Resources (DERs) in the service territories of California's Pacific Gas and Electric (PG\&E) and Southern California Edison (SCE). Results demonstrate that a high correlation exists between race and grid limits in these regions, such that in Black--identifying and disadvantaged communities, hosting capacity \cite{hosting_cap} for DERs drastically decreases \cite{Brockway}, hindering EV adoption in these neighborhoods. Although, existing literature caters to the socio-demographic and census block group--based analysis of equitable distribution of EV charging infrastructure, it mainly focuses on qualitative discussions \cite{Hardman} or data-driven analysis of isolated socio-economic factors \cite{Brockway, Canepa, Hsu}. To the best of authors' knowledge, a systematic analysis of correlations between socio-demographic features and their mutual effect on the access and affordability of EV charging infrastructure is missing. Moreover, the analyses in this paper are carried out with a zip code--level granularity, which is aptly suited for an urban justice setting.

In this paper, our hypothesis is that the distribution of EV charging stations is closely related to the inter-dependent socio-demographic features of population. We further hypothesize that features of the local transportation landscape may also be related to the distribution of EV charging stations. For instance, owing to a high influx of traffic, zip codes with high concentrations of major roadways may be more desirable locations for charging stations. Hence, the identification and quantification of such features is of paramount importance. We consider socio-demographic features like population size, median household income, poverty rate, and racial makeup of population, and transportation features like presence of highways and number of highways in each zip code. These features serve as markers to the current imbalances in the accessibility and affordability levels of EV charging stations in the society \cite{justice}. In this paper, we do not seek to furnish causal claims, however, identify correlations that exist in data so that targeted policy interventions can be designed to facilitate an equitable roll-out of EV charging infrastructure.

\section{Data}
\label{Sec:Data}
To determine the socio-demographic and transportation factors affecting the distribution of EV charging stations in NYC, we use the publicly available Alternative Fuel Station Locator dataset from Alternative Fuel Data Centre at the US Department of Energy \cite{EV_data}. This dataset provides a current accounting of the types and locations of all alternative fuel stations in NYC. For this analysis, we include only electric charging stations and exclude those providing other alternative fuels, like biodiesel, compressed natural gas (CNG), or liquefied natural gas (LNG). Each data point in the EV charging station dataset corresponds to one station, irrespective of the number of EV service equipment ports (charging outlets) and the type of connectors. The data comprises charging stations operated by major EV charging companies in the US, including Blink, ChargePoint, Electrify America, EVgo, FLO, Greenlots, OpConnect, Tesla, SemaConnect, and Webasto, however, does not include residential EV charging locations. Similarly, we obtain the demographic data of NYC from the American Community Survey (ACS) \cite{dem_data}. We use ACS 1--year estimates data profiles for 2019, which includes features of median household income, poverty rate, and population percentage of different racial groups by zip code. Using the NYS Streets data from the New York State (NYS) GIS Offices, we obtain the routes and spatial information of major roadways, which include interstates, interstate connections, state touring routes and connectors, state 900 routes, US highways, and US highway business routes and connectors \cite{highway_data}, in NYC. The transportation information is then mapped to individual zip codes of NYC, shown in Fig.~\ref{NYC}, using the zip code boundaries  dataset from the Department of Information Technology \& Telecommunications \cite{zipcode_data}. The socio-demographic and transportation data used in this paper is available in \cite{data}. 

\section{Methods} \label{methods}
\begin{figure}[!t]
\centering
\includegraphics[width=2.8in, height = 2.8in]{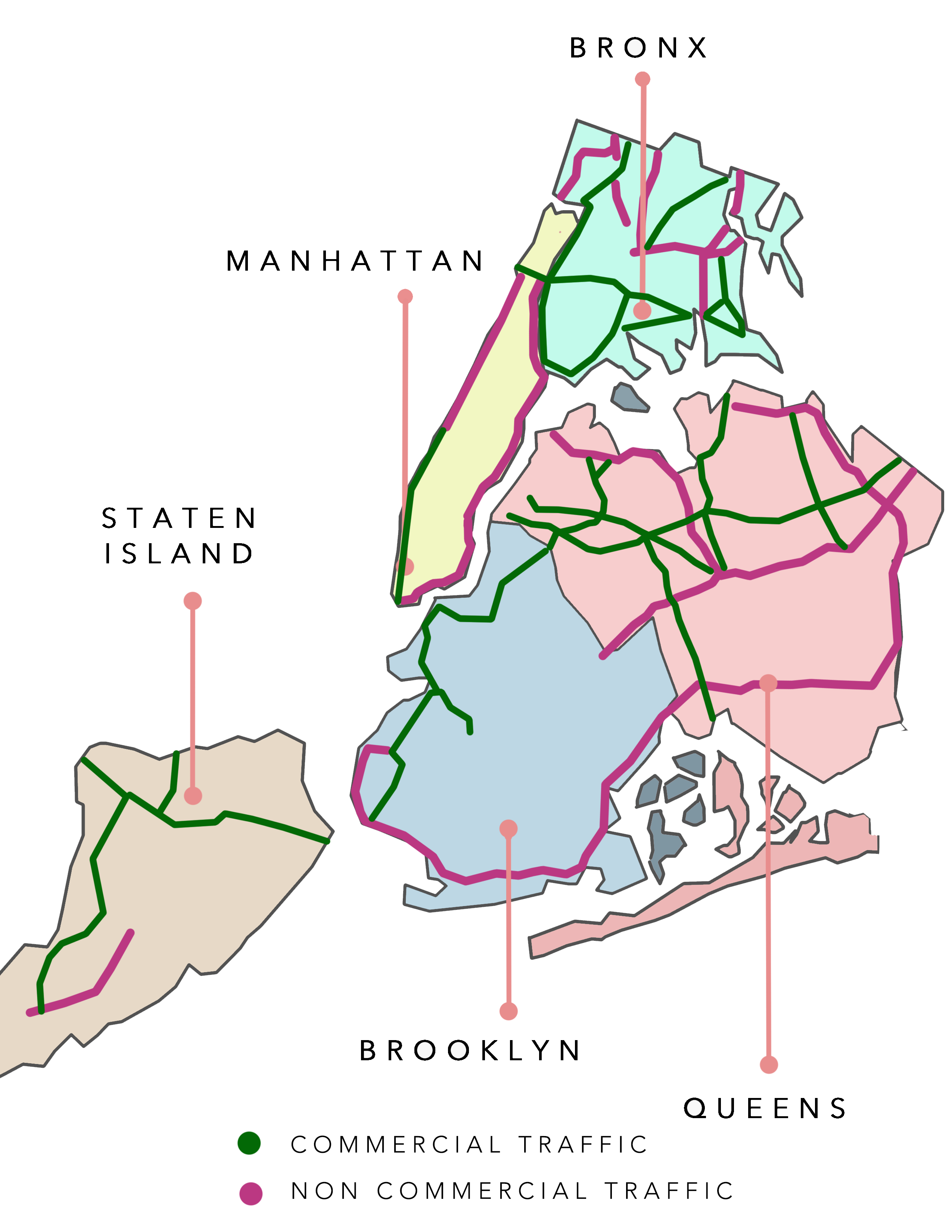}
\caption{\small{Map showing the five constituent boroughs of NYC, along with the commercial and non-commercial traffic routes (highways).}}
\vspace{-20pt}
\label{NYC}
\end{figure}
We perform correlation analyses, in Section~\ref{Corr}, to identify features in the demographic and highway datasets that impact the distribution of EV charging stations in NYC. To this end, we define the following two target features:
\begin{enumerate}
\item A binary variable representing the presence of at least one EV charging station
in a particular zip code
\item The total number of EV charging stations in each zip code
\end{enumerate}

Moreover, we test the hypotheses in Section~\ref{intro} by defining two groups in our dataset, such that all zip codes with at least one EV charging station constitute one group whereas the remaining zip codes constitute the other group. Across the five boroughs of NYC, shown in Fig.~\ref{NYC}, there are 180 zip codes with accompanying demographic data from the ACS \cite{dem_data}. 100 of these zip codes have at least one EV charging station and 80 have no EV charging stations. Hence, group 1 in our dataset comprises 100 data points, whereas group 2 contains 80 data points. Owing to a normal distribution of EV charging station data and an almost equal sample size of the two data groups, we use \textit{t}-test as a hypothesis testing tool, in Section~\ref{hypo}, to compare the average values of the two groups \cite{t-test-ref}. Our null hypothesis assumes that means of the two data groups are equal, i.e., there are no statistical differences between the two groups. The null hypothesis implies that the socio-demographic features do not affect the presence of EV charging stations in zip codes, rendering the two groups statistically identical.        
We use both \textit{p}-value and \textit{t}-value to assess the likelihood to reject the null-hypothesis. In this case, we use the \textit{p}-value $\leq 0.05$ as statistically significant \cite{p-value}, indicating a strong evidence against the null hypothesis. On the contrary, for significance level ($\alpha$) = 0.05, \textit{t}-value is significant if $\mid \textit{t} \mid \geq 1.96$.

While correlations can be identified between individual demographic features and
distribution of EV charging stations in NYC, the inter-dependency of these features
cannot be ruled out. For example, the median household income of a particular zip
code may or may not be related to the racial makeup of its population. Therefore, we
analyze the dependency between socio-demographic features using conditional analysis in Section~\ref{conditional}. The number of EV charging stations in each zip code is analyzed as a function of the percentage of population
identifying as white or non--white, conditioned on a threshold income.
We choose this threshold as the annual median income of NYC for 2015--2019, which is estimated to be \$64,000 \cite{dem_data}.

\section{Case Study}

\begin{figure}[!t]
\centering
\includegraphics[width=3.3in, height = 2.5in]{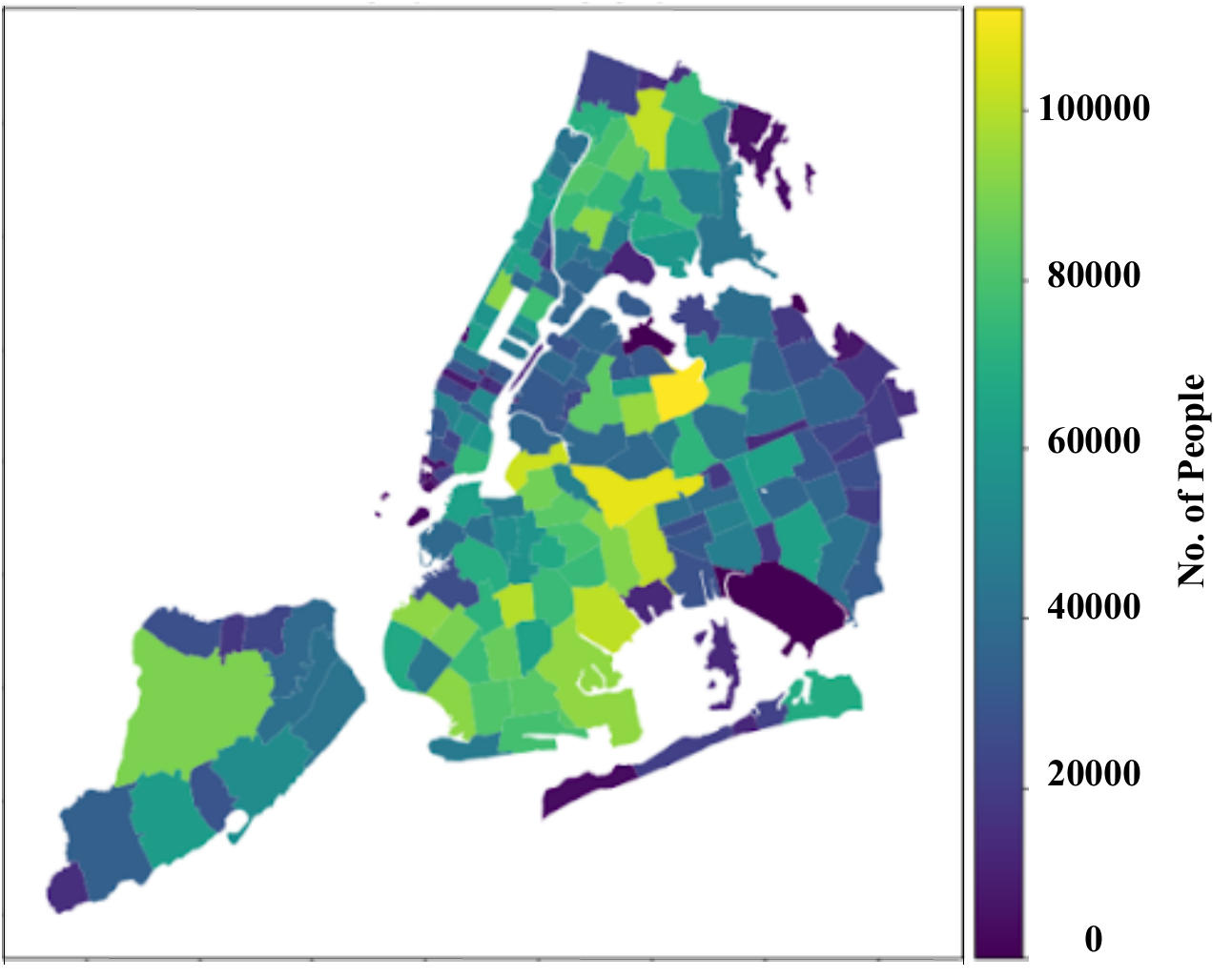}
\caption{\small{Heat map depicting the zip code--level population density in NYC, based on data in \cite{dem_data}.}}
\vspace{-20pt}
\label{population}
\end{figure}

The case study focuses on NYC, with Fig.~\ref{NYC} showing its five constituent boroughs -- Manhattan, Bronx, Queens, Brooklyn, and Staten Island -- in different colors. In this section, we use the demographic and EV charging station data with a zip code resolution, described in Section~\ref{Sec:Data}, to retrieve correlations between different demographic features and the distribution of EV charging stations across NYC. Hypothesis testing is performed between the set of zip codes with at least one EV charging station and the set of zip codes with no EV charging stations to ascertain the features that distinguish these two sets. Finally, to understand the inter-dependencies of demographic features, we use conditional analysis to see the distribution of EV charging stations in different zip codes of NYC. 

\begin{figure}[!t]
\centering
\includegraphics[width=3.1in, height = 2.5in]{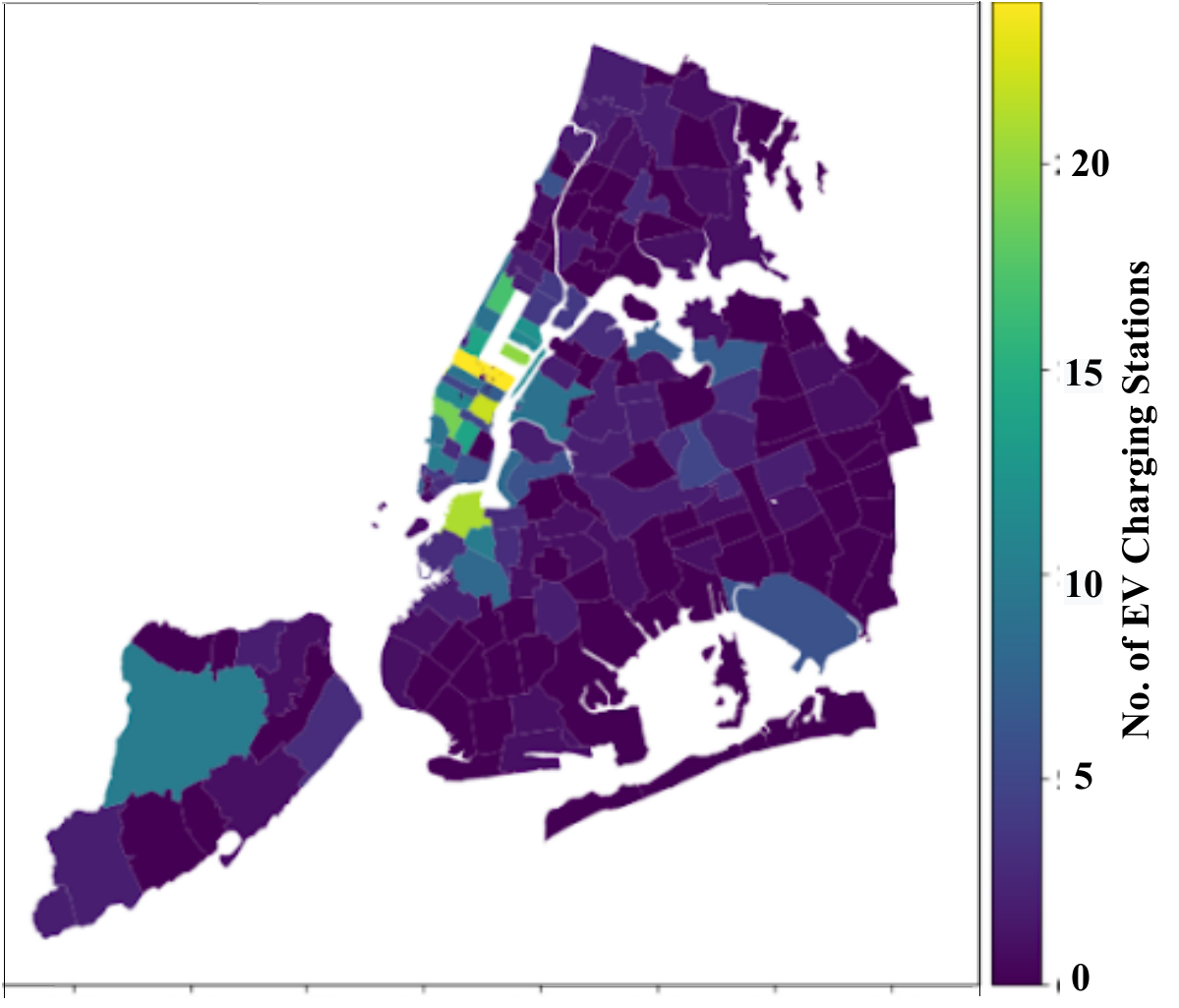}
\caption{\small{Zip code--level distribution of EV charging stations in NYC, based on data in \cite{EV_data}.}}
\vspace{-10pt}
\label{dist_EV}
\end{figure}

Using data from \cite{dem_data}, we show the population density in each zip code of NYC in Fig.~\ref{population}. We note that the population density in NYC is neither uniform across zip codes nor across the five boroughs. Using Figs.~\ref{NYC} and \ref{population}, we observe that Brooklyn has the highest population density (shown in yellow in Fig.~\ref{population}) whereas average population density is very low in Staten Island and Manhattan. Similarly, using the EV charging station data in \cite{EV_data}, we show the distribution of EV charging stations in NYC on a zip code level, in Fig.~\ref{dist_EV}. We note that the distribution of EV charging stations in heavily non-uniform among different boroughs such that the maximum number of charging stations are concentrated in Manhattan, whereas Bronx, Queens, and Brooklyn have little to no charging stations in most of the zip codes.

As an initial hypothesis, we compare the trends in population density in zip codes and boroughs to the associated distribution of EV charging stations. Comparing Figs.~\ref{population} and \ref{dist_EV}, we note that there exists a huge disparity between the distribution of EV charging stations and population density in NYC. While geographical areas (multiple contiguous zip codes) in Brooklyn have some of the highest population densities, the same areas have very few EV charging stations. Similarly, most of the EV charging stations are concentrated in Manhattan, where population density is one of the lowest. This indicates that population density is not a good indicator for the distribution of EV
charging stations, underscoring inaccessibility of residential EV charging infrastructure as an acute barrier to EV adoption. Thus, further analysis of socio-demographic data is required to capture
features that determine the presence and accessibility of EV charging stations. In
the following sections, we analyze the demographic characteristics that correlate with the development and allocation
of EV charging stations in NYC. 

\subsection{Correlation Analysis}\label{Corr}
\begin{figure}[!t]
\centering
\includegraphics[width=3.1in, height = 2.5in]{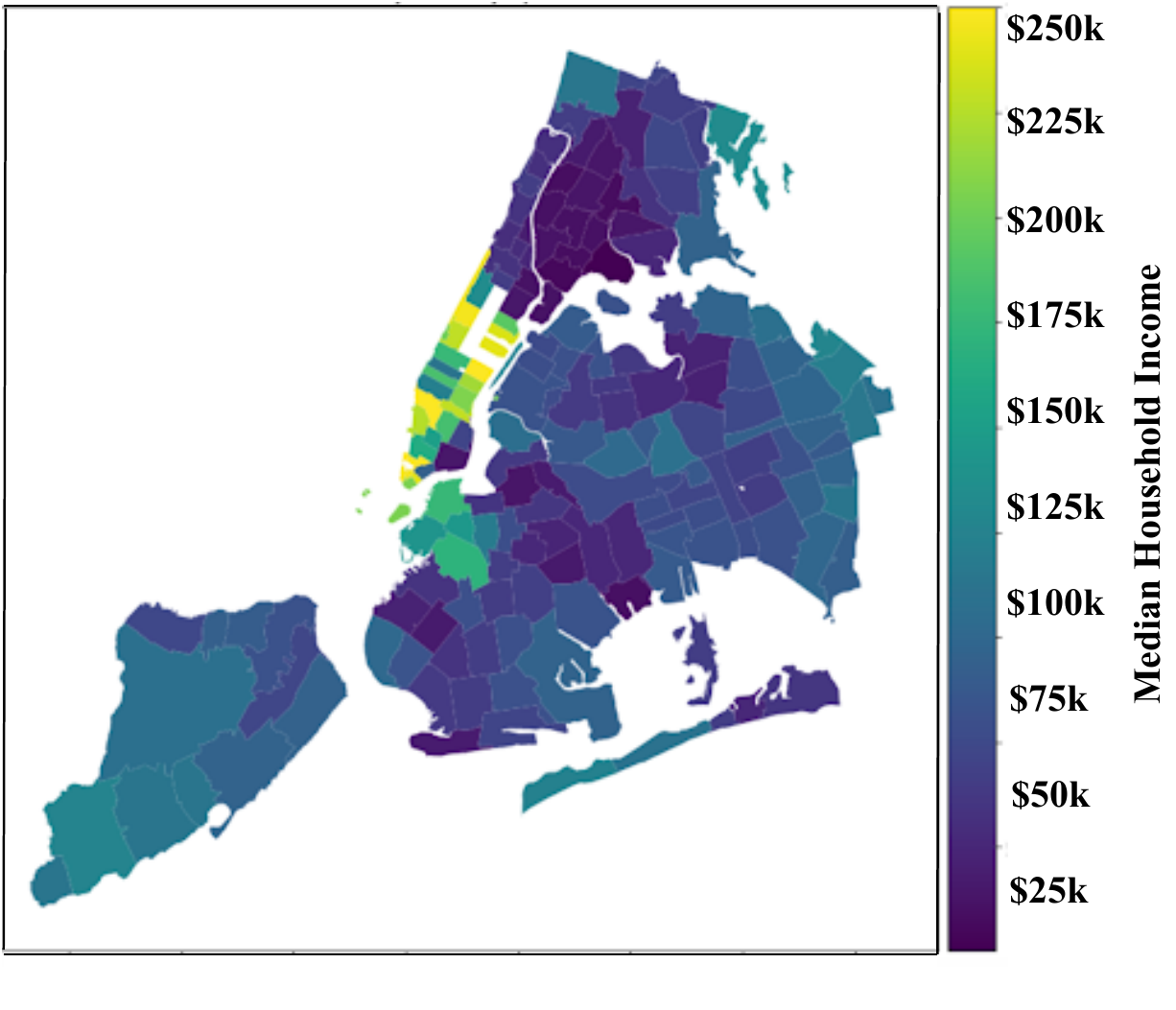}
\caption{\small{Distribution of household income in NYC by zip code, based on data in \cite{dem_data}.}}
\vspace{-20pt}
\label{income}
\end{figure}

\begin{figure}[!t]
\centering
\includegraphics[width=3.1in, height = 2.5in]{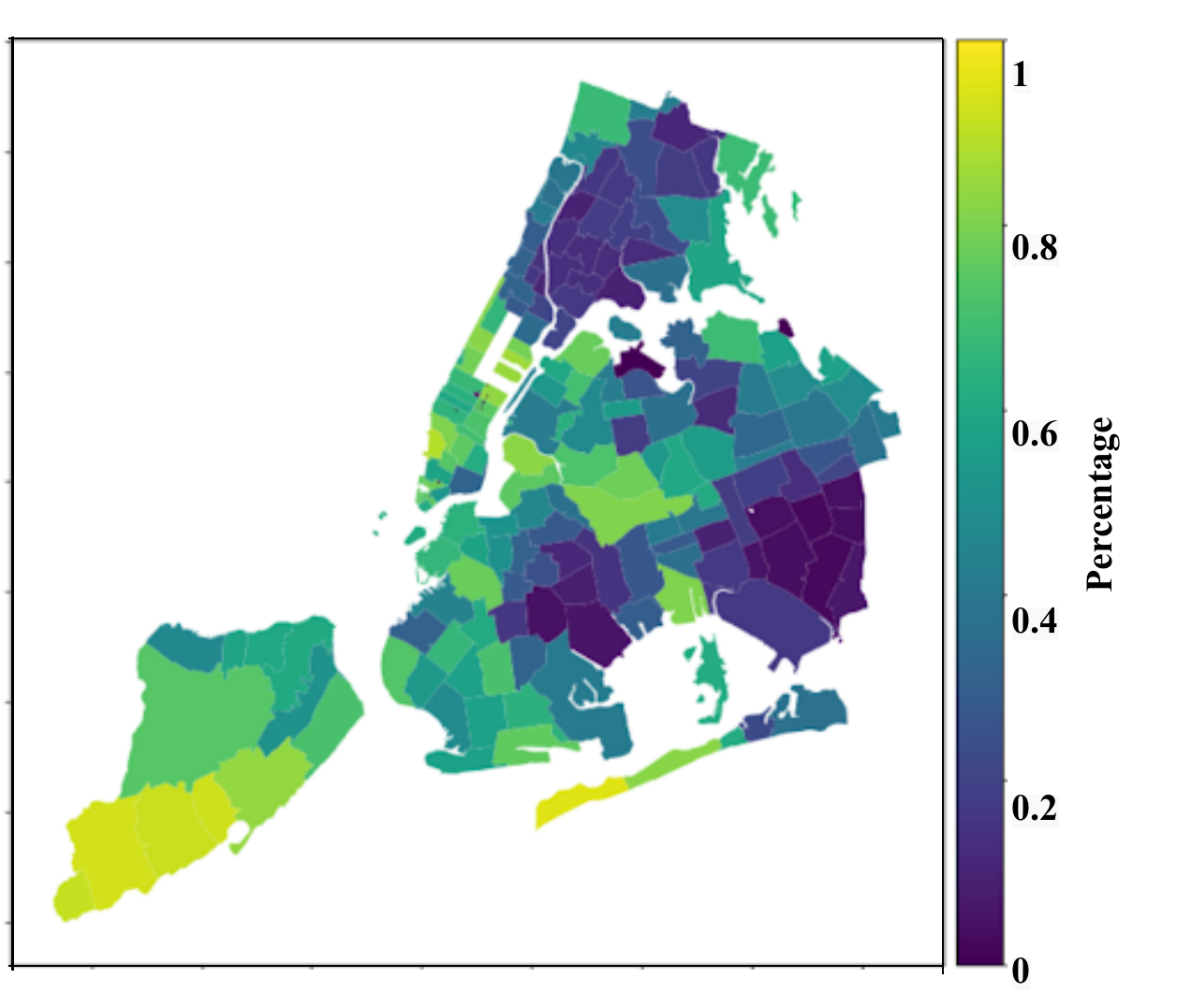}
\caption{\small{Percentage of White--identifying population in NYC by zip code, based on data in \cite{dem_data}.}}
\vspace{-15pt}
\label{white}
\end{figure}

\vspace{-20pt}
\begin{center}
\begin{table}[!t]
\caption{Correlation between demographic/transportation features and the target features of whether an EV charging station is present and the number of EV charging staions in a zip code.}
\centering
\begin{tabular}{ >{\centering}p{4.2cm} | >{\centering}p{1.8cm} | >{\centering\arraybackslash}p{1.8cm}}
\hline
\hline
Demographic Feature & Station Present & No. of Stations   \\
\hline 
Median household income & 0.45 & 0.58  \\
White--identifying population (\%) & 0.43 & 0.43  \\
Highway present & 0.32 & 0.23  \\
Asian--identifying population (\%) & 0.24 & 0.16 \\
Highway count & 0.2 & 0.07 \\
Poverty rate & 0.2 & 0.01 \\
Hispanic--identifying population (\%) & 0.18 & -0.06 \\
Black--identifying population (\%) & -0.02 & -0.14 \\
\hline
\end{tabular}
\vspace{-10pt}
\label{correlation}
\end{table}
\end{center}
Table~\ref{correlation} shows the results of the correlation analysis. Median household income and
percentage of White--identifying population in a zip code show the highest positive correlation
with the presence of at least one EV charging station and the number of stations present in that
zip code. Hence, higher the median income of a given zip code, the higher the probability
that at least one EV charging station will be present in that zip code. Similarly, a higher
median income also implies a higher number of EV charging stations in a zip code. The
same pattern holds when we compare the relationship between the percentage of White--identifying population in each zip code and the two aforementioned target features. This trend can be visualized using Fig.~\ref{income}, which shows that zip codes in Manhattan have the highest median household income. These zip codes also have the highest number of EV charging stations, as given by a correlation coefficient of 0.58 in Table~\ref{correlation}. Similarly, Bronx has
the lowest median income and the lowest number of EV charging stations.
We note in Table~\ref{correlation}, that the correlation coefficient between percentage of White--identifying
population and the presence of EV charging stations is smaller as compared to the one
between median household income and presence of EV charging stations. This can be explained using  Fig.~\ref{white}, where some zip codes in Manhattan have a notably high percentage of White--identifying
population (and a high median household income) and a high number of EV charging
stations. However, the trend does not hold in Staten Island and large portions of Brooklyn, which
have similarly high percentages of White--identifying population but a very low penetration of EV charging stations. We also observe a weak positive correlation of 0.32 between the presence of highways in a zip code and the number of EV charging stations present in that zip code.

Hence, although the percentage of White--identifying population is a good indicator for predicting the distribution of EV charging stations across zip codes in NYC, certain regions in Staten Island and Brooklyn do not follow this correlation. However,  median household income explains the distribution of EV charging stations throughout NYC. 

\subsection{Hypothesis Testing} \label{hypo}
Based on the two data groups defined in Section~\ref{methods}, we perform a two-sample \textit{t}-test between these two populations for different demographic and transportation features to determine (dis)similarities between the two groups. Table~\ref{T-test} shows the results of this analysis along with the mean and median values of the features for each group. The mean values for median household income
and percentage of White--identifying population for zip codes with at least one charging station
are significantly higher than for zip codes without any charging station. Meanwhile,
the mean value for percentage of Black--identifying population is significantly lower in zip codes
with at least one charging station versus those without any.
\vspace{-20pt}
\begin{center}
\begin{table*}[!t]
\caption{Results of hypothesis testing between the group of zip codes with and without EV charging stations.}
\centering
\begin{tabular}{ >{\centering}p{5.5cm} | >{\centering}p{1.5cm} | >{\centering}p{1.5cm} | >{\centering}p{1.5cm} | >{\centering}p{1.5cm} | >{\centering}p{1.5cm} | >{\centering\arraybackslash}p{2cm}}
\hline
\hline
\multirow{2}{*}{Demographic Feature}& \multicolumn{2}{c|}{Mean} & \multicolumn{2}{c|}{Median} & \textit{t}-stat & \textit{p}-value   \\
\cline{2-5}
& Station & No Station & Station & No Station &  &\\
\hline 
Median household income & 112,000 & 78,300 & 87,242 & 75,471 & 4.02 & .000087 \\
Poverty rate & 15.39  & 17.29 & 13.18 & 13.78 & -1.21 & .23   \\
Percentage of White--identifying population & 51.79 & 39.08 & 56.37 & 37.01 &  3.36 & .00094\\
Percentage of Black--identifying population & 16.79 & 29.12 & 7.26 & 22.05 & -3.37 & .00094\\
Percentage of Hispanic--identifying population & 25.06 & 27.79 & 17.05 & 19.94 & -0.93 & .35 \\
Percentage of Asian--identifying population & 15.81 & 13.95 & 10.48 & 7.25 & 0.87 &  .39 \\
Highway count & 1.79 & 1.20 & 2 & 0.5 & -0.83 & .41 \\
Highway present & 0.81 & 0.50 & 1 & 0.5 & 5.15 &  .00000056\\
\hline
\end{tabular}
\label{T-test}
\end{table*}
\end{center}

Based on the results of two-sample \textit{t}-test, reported in  Table~\ref{T-test}, we observe that median household income, percentage of White--identifying population, percentage of Black--identifying population, and presence of highways in zip codes offer significant \textit{t}-values and \textit{p}-values. Hence, using the \textit{p}-values, we can reject the null hypothesis for these features, concluding that the means of the two groups of zip codes are not equal, indicating that there are significant statistical differences between the two groups.   Similarly, the same features have significant \textit{t}-values, indicating that the two groups of zip codes are dissimilar from one another. 

Our analysis concludes that presence of highways is the most significant feature that distinguishes the two data groups, followed by median household income, and percentages of White-- and Black--identifying population. The results complement the trends observed in correlation analysis, and offer critical insights into demographic features that determine the distribution of EV charging stations across NYC.

\begin{figure}[!t]
\centering
\includegraphics[width=3.1in, height = 2.5in]{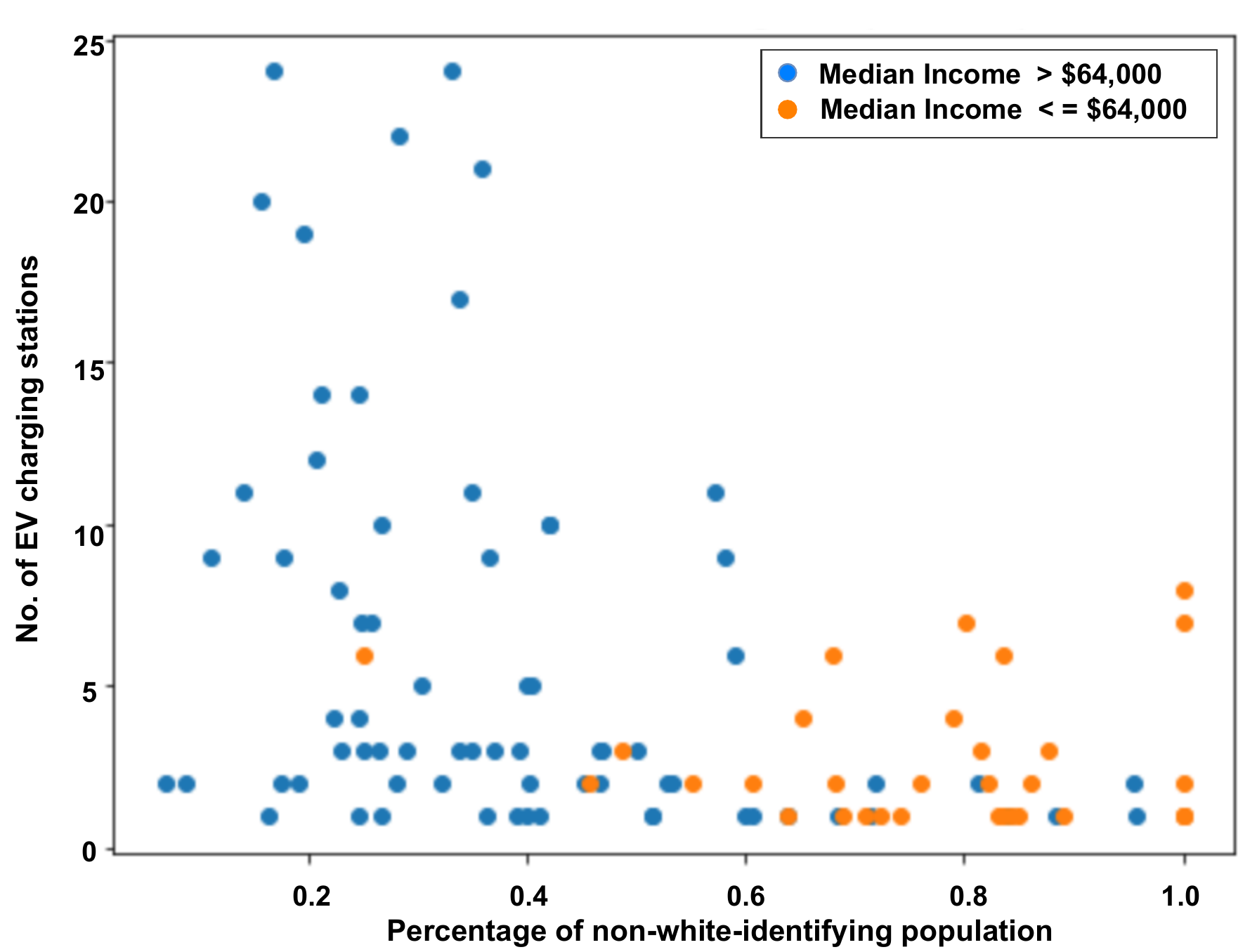}
\caption{\small{Conditional analysis of percentage of non-White-identifying population and the number of EV charging stations in a zip code, conditioned on median income of NYC. }}
\label{comp_nonwhite}
\vspace{-20pt}
\end{figure}

\begin{figure}[!t]
\centering
\includegraphics[width=2.9in, height = 2.5in]{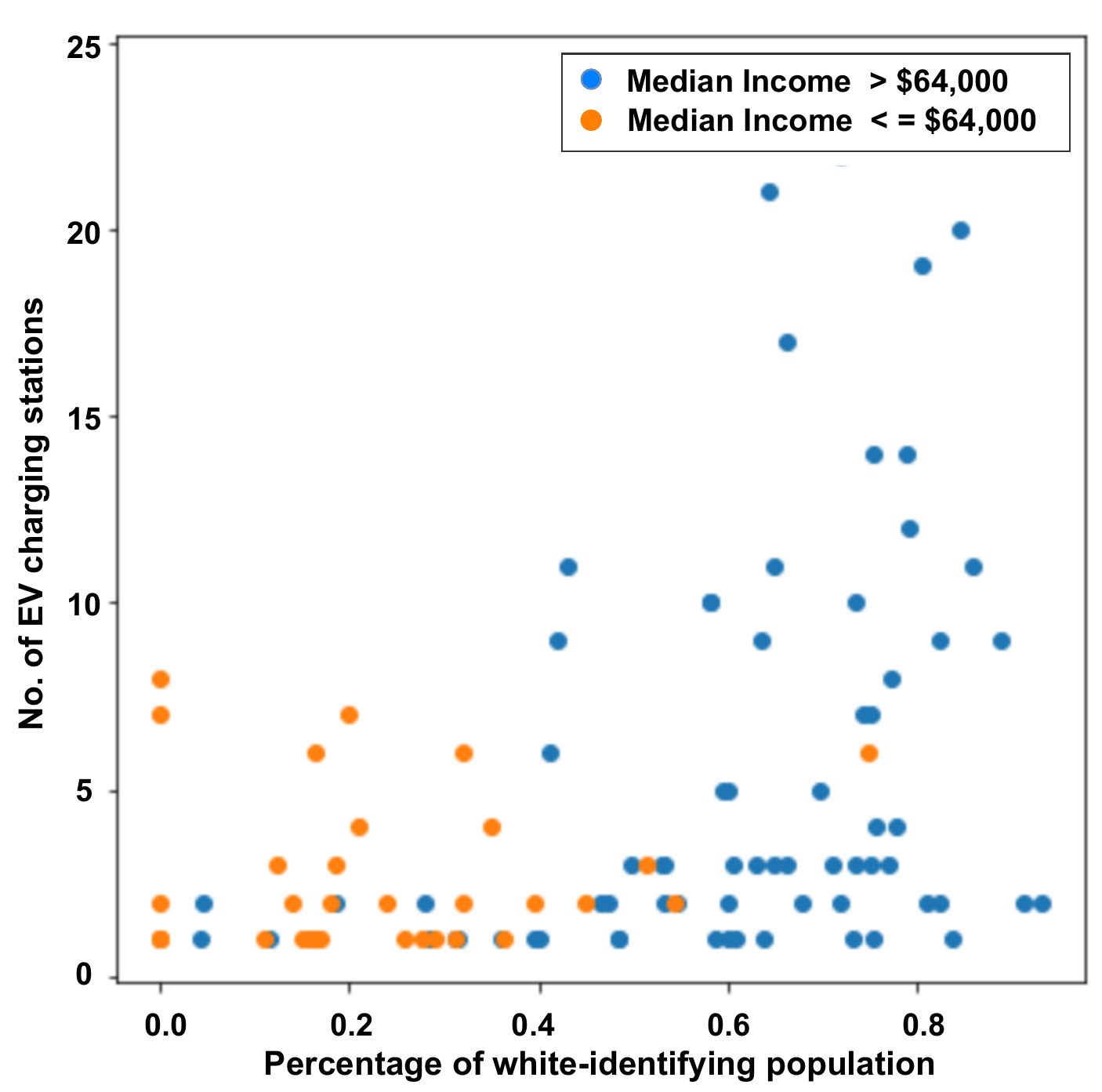}
\caption{\small{Conditional analysis of percentage of White-identifying population and the number of EV charging stations in a zip code, conditioned on median income of NYC.}}
\vspace{-25pt}
\label{comp_white}
\end{figure}

\subsection{Conditional Analysis} \label{conditional}
Figs.~\ref{comp_nonwhite} and \ref{comp_white} show the results of conditional analysis where we identify the existence of conditional trends. We see that in zip codes with median income greater than \$64,000,
there is a clear positive correlation between the number of EV charging stations and the percentage of White--identifying population. Similarly, in zip codes with median income less than
or equal to \$64,000, a negative correlation exists between the number of EV charging
stations and the percentage of non-White--identifying population. These results show that although the distribution of EV charging stations in NYC can be correlated to individual features of dataset, a better understanding of EV charging infrastructure is acquired by intersecting economic and racial profiles of population. 

\section{Conclusion}
This paper analyzes the socio-demographic and transportation features that affect the distribution of EV charging stations in NYC. Based on correlation analysis, hypothesis testing, and conditional analysis, our results demonstrate that the availability and affordability of EV charging stations in
NYC are not determined by the population density, but are correlated with the median household income, percentage of White--identifying population, and presence of highways in a zip code. The current distribution is heavily skewed against low--income, Black--identifying, and disadvantaged neighborhoods. 
The existing inequities in EV infrastructure, coupled with underdeveloped electricity, transportation, or communication infrastructures in vulnerable communities, may result in biased and inequitable siting decisions for future EV charging infrastructure roll--out. Therefore,  justice--centric policy frameworks are imperative to ensure large-scale and equitable EV adoption.


\bibliographystyle{IEEEtran}
\bibliography{references}{}

\end{document}